\documentclass[reprint,superscriptaddress,amsmath,amssymb,aps,prl]{revtex4-1}
\usepackage{color, amssymb, graphicx, amsfonts,epstopdf }
\usepackage{multirow,amssymb,amsbsy,amsmath,epstopdf}
\usepackage{hyperref}
\usepackage{graphicx}
\usepackage{verbatim}
\usepackage{bm}
\usepackage{bbold}

\newcommand{\ket}[1]{\ensuremath{\left|#1\right\rangle}}

\begin{document}

\title{Photonic realization of erasure-based nonlocal measurements}
\author{Wei-Wei Pan}
\author{Xiao-Ye Xu}
\affiliation{CAS Key Laboratory of Quantum Information, University of Science and Technology of China, Hefei 230026, People's Republic of China}
\affiliation{CAS Center For Excellence in Quantum Information and Quantum Physics, University of Science and Technology of China, Hefei 230026, People's Republic of China}
\author{Eliahu Cohen}
\email{eliahu.cohen@biu.ac.il}
\affiliation{Faculty of Engineering and the Institute of Nanotechnology and Advanced
Materials, Bar Ilan University, Ramat Gan 5290002, Israel}
\author{Qin-Qin Wang}
\author{Zhe Chen}
\author{Munsif Jan}
\author{Yong-Jian Han}
\author{Chuan-Feng Li}
\email{cfli@ustc.edu.cn}
\author{Guang-Can Guo}
\affiliation{CAS Key Laboratory of Quantum Information, University of Science and Technology of China, Hefei 230026, People's Republic of China}
\affiliation{CAS Center For Excellence in Quantum Information and Quantum Physics, University of Science and Technology of China, Hefei 230026, People's Republic of China}

\begin{abstract}
Relativity theory severely restricts the ability to perform nonlocal measurements in quantum mechanics. Studying such nonlocal schemes may thus reveal insights regarding the relations between these two fundamental theories. Therefore, for the last several decades, nonlocal measurements have stimulated considerable interest. 
%
However, the experimental implementation of nonlocal measurements imposes profound restrictions due to the fact that the interaction Hamiltonian cannot contain, in general, nonlocal observables such as the product of local observables belonging to different particles at spacelike-separated regions. 
In this work, we experimentally realize a scheme for nonlocal measurements with the aid of probabilistic quantum erasure. 
We apply this scheme to the tasks of performing high accuracy nonlocal measurements of the parity, as well as measurements in the Bell basis, which do not necessitate classical communication between the parties. Unlike other techniques, the nonlocal measurement outcomes are available locally (upon successful postselection).
The state reconstructed via performing quantum tomography on the system after the nonlocal measurement indicates the success of the scheme in retrieving nonlocal information while erasing any local data previously acquired by the parties. 
This measurement scheme allows to realize any controlled-controlled-gate with any coupling strength. Hence our results are expected to have conceptual and practical applications to quantum communication and quantum computation.
\end{abstract}

\maketitle

\section{Introduction}
Quantum nonlocality\,\cite{Einstein1935,Brunner2014} is intriguing in that it allows, on the one hand, to establish correlations and achieve tasks which are classically impossible, but on the other hand, it does not violate relativistic causality.

In the following work, we focus on a specific scenario: two parties, Alice and Bob, are located in remote positions (possibly space-like-separated), but they wish to perform a joint quantum measurement of their nonlocal system. This goal has motivated the study of quantum nonlocal measurements, which lie at the interface of quantum mechanics and relativity theory\,\cite{Landau1931,Aharonov1981,Aharonov1986,Popescu1994,Beckman2001,Groisman2002,Vaidman2003,Harrow2011}. The former theory provides Alice and Bob with tools, such as quantum entanglement, for accomplishing the task, while the latter sets limitations on the causal relations between them. Indeed, many nonlocal observables cannot be instantaneously measured in a non-demolition projective measurement\,\cite{Popescu1994,Clark2010}, while others can be destructively measured\,\cite{Groisman2002,Vaidman2003} with a weakened coupling strength and a suboptimal ratio of information/disturbance\,\cite{Brodutch2016}.

\begin{figure}
   \centering
   \includegraphics[width =0.85\linewidth]{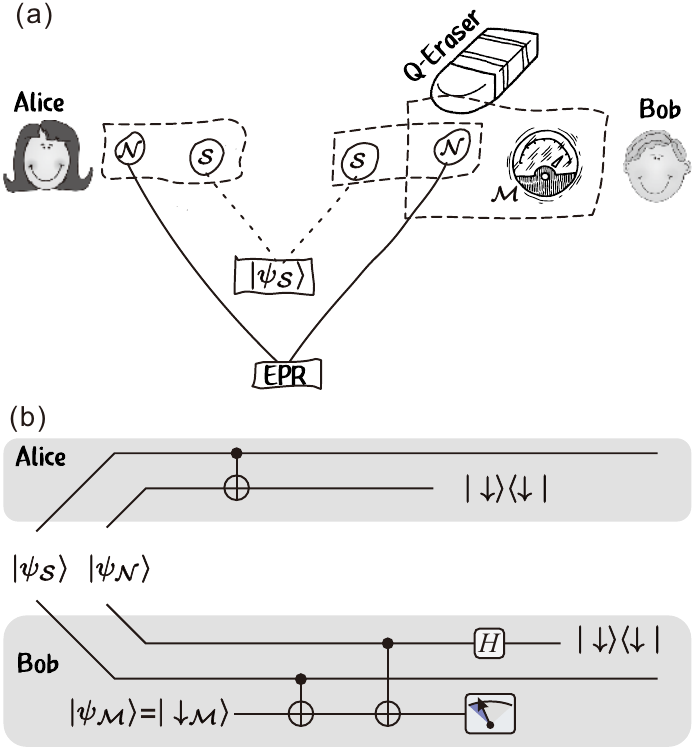}
   \caption{{\bf Theoretical scheme} Conceptual diagram (a) and the corresponding quantum logic circuit (b) for the erasure-based nonlocal measurement. Alice and Bob share two EPR pairs, one of which acts as the system $\mathcal{S}$ to be measured (this state could in fact be any entangled state, or even a product) and the other acts as the measurement apparatus $\mathcal{N}$. On Bob's side, a meter $\mathcal{M}$ is introduced. After performing von Neumann coupling to $\mathcal{N}$ and erasing the information recorded by $\mathcal{N}$, the complete information of the system is then captured by $\mathcal{M}$. All the measurement couplings are realized by C-NOT gates, similarly to the proposal in\,\cite{Brodutch2016}. The quantum erasure step is implemented with a Hadamard gate in the qubit scenario, as depicted in (b).}
   \label{fig:logic}
\end{figure}

Performing nonlocal measurement is prevalent in quantum mechanics, see e.g.\,\cite{Bennett1999,Beckman2001,Groisman2002,Paneru2017,Aharonov2017, Paraoanu2018}, and is actually a crucial step in some schemes for quantum information processing, e.g. error correction\,\cite{Gottesman1997}, device-independent quantum key distribution\,\cite{Matthew2009,Stefano2009,Lim2013}, and realization of multipartite gates in general.

Several protocols have been suggested over the years for realizing such measurements in the strong\,\cite{Aharonov1981,Aharonov1986,Beckman2001,Groisman2002,Vaidman2003,Harrow2011} and weak\,\cite{Resch2004,Brodutch2009,Kedem2010} coupling regimes. Recently, Brodutch and Cohen have proposed a new method\,\cite{Brodutch2016}, which is based on quantum teleportation\,\cite{Ben1993,Bou1997,Boschi1998} and quantum erasure\,\cite{Scully1982,Herzog1995} - Alice performs a local measurement of her system and then teleports the outcome to Bob. Bob then couples the resulting system to his local measurement device and makes a measurement, followed by a probabilistic erasure of the local superfluous information. When Bob's postselection is successful, his measurement device shifts by an amount proportional to an eigenvalue of the nonlocal observable (thereby being linear in the coupling strength), just like in the case of von Neumann measurement of this observable. A failure in the postselection stage corresponds to a non-trivial unitary evolution of the state between pre- and postselection which can be later corrected given classical communication between Alice and Bob. This protocol was shown in\,\cite{Brodutch2016} to be more versatile than other methods, allowing to measure a wider class of nonlocal, multipartite observables with a variable coupling strength. Furthermore, it can be used for implementing any controlled-controlled-unitary operation, as well as measurements of non-Hermitian operators\,\cite{Brodutch2017}. However, this erasure method ought to be probabilistic\,\cite{Brodutch2016} in order to preserve relativistic causality, thus exemplifying the necessity of quantum uncertainty in nonlocal scenarios\,\cite{Aharonov2017,Carmi2018}.

\begin{figure*}
    \centering
    \includegraphics[width = 0.9\textwidth]{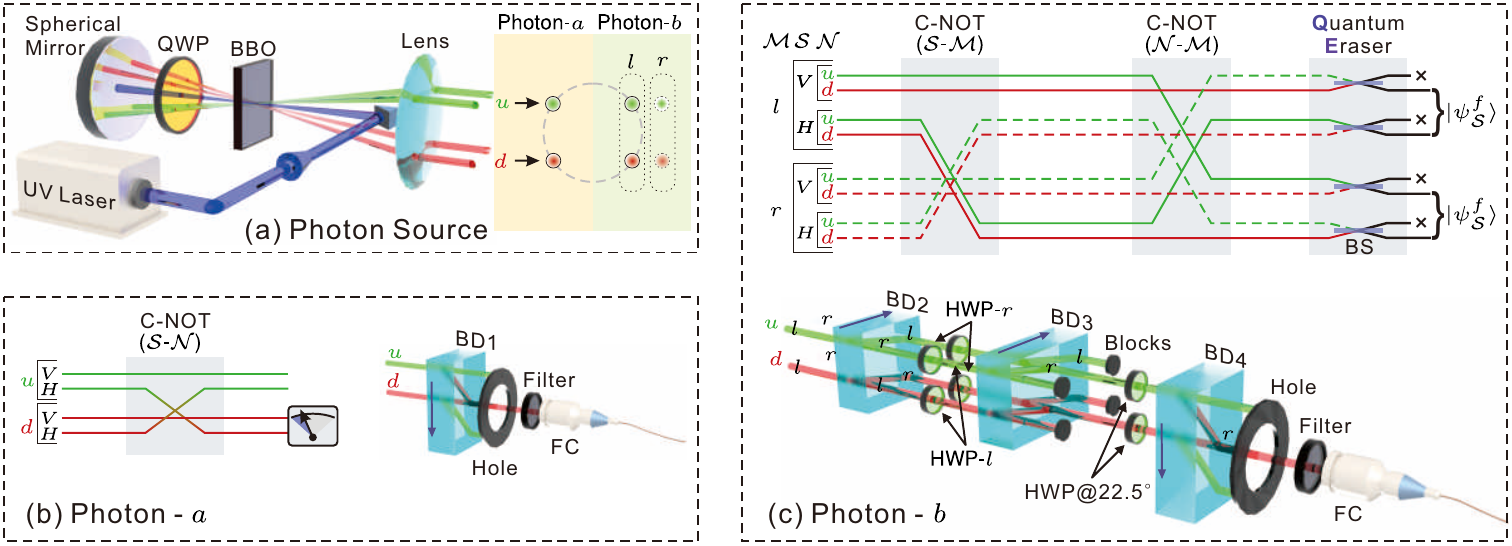}
    \caption{\textbf{Experimental setup.} (a) shows the photon source prepared in our experiment, we adopt the structure used in Ref.\,\cite{Ciampini2016}. A vertically polarized ultraviolet (UV) laser (wavelength centered at 406.7\,nm) pumps a $\beta$-BaB$_2$O$_4$ (BBO) crystal to generate photon pairs (selected by spectrum filters with center at 813.4\,nm and bandwidth 3\,nm ) in a polarization-momentum hyper-entangled state. We collect four points in the entanglement ring, and denote the left part as photon-$A$ and the right part as photon-$B$ respectively. For photon-$B$, we introduce another degree of freedom (left and right modes) to act as the meter $\mathcal{M}$. (b) shows the operations on Alice's side. Left part schematically illustrates the quantum circuit. A beam displacer (BD1) with its optical axis cut in the vertical plane, which transmits vertically polarized light straightly and displaces horizontally polarized light in the vertical plane, is used to realize the C-NOT gate (polarization control path). The hole helps us to collect the photons in path-$d$ by a fiber coupler (FC) after a spectrum filter. The operations by Bob are shown in (c) which includes the schematic diagram in the upper panel and experimental realizations in the lower panel. The C-NOT gate between system and meter is realized by BD2 (horizontally cut), where vertically-polarized photons propagate directly and horizontally-polarized photons are shifted in the horizontal plane. Four half wave plates (HWPs) collaborate with another horizontally cut BD3 to arrange that the two $l$ ($r$) beams (green and red) lie in the same vertical plane. Two HWPs with their optical axes oriented at $22.5^\circ$ collaborate with a vertically cut BD4 to realize the quantum eraser, which erases the quantum information encoded in the ancilla $\mathcal{N}$. QWP stands for quarter wave plate at 813.4\,nm.}
    \label{fig:setup}
\end{figure*}

An experimental scheme for realizing this erasure-based protocol was proposed in\,\cite{Wu2016}, which requires a challenging technique, i.e., nonlinear interaction between single photons. Here, based on the method for implementing quantum C-NOT gates between different photonic degrees of freedom\,\cite{Kim2003,Fiorentino2004}, we opted for a different, more feasible realization which is presented below. 

The basic idea of our erasure-based nonlocal measurement can be seen from the conceptual diagram presented in Fig.\,\ref{fig:logic}. Let us consider the task of performing a nonlocal measurement over a system composed of the two-qubit state $\ket{\psi_\mathcal{S}} = \sum_{\mu,\nu}\psi_{\mu,\nu}\ket{\mu\nu}$, where $\mu,~\nu\in\{\uparrow,\downarrow\}$ and $\psi_{\mu,\nu}$ are the corresponding complex amplitudes. We now need an entangled ancilla $\mathcal{N}$ initialized in an EPR state $\ket{\psi_\mathcal{N}} = \frac{1}{\sqrt{2}}(\ket{\uparrow\downarrow} + \ket{\downarrow\uparrow})$, where one particle is held by Alice and the other is held by Bob. First, Alice performs a local measurement on her side and gets the outcome $\sigma_z = -1$. Then Bob couples the meter $\mathcal{M}$ (initialized in its ground state) to both the system $\mathcal{S}$ and his part of the ancilla $\mathcal{N}$. After erasing the information contained in $\mathcal{N}$, the global information of the system (for example, the parity $\mathcal{P}$) is then given by the meter state $\ket{\psi_\mathcal{M}}$. All measurement couplings here are of the von Neumann type, which can be realized via C-NOT gates, as shown in Fig.\,\ref{fig:logic}(b). The quantum erasure of the ancillary state is implemented with a Hadamard gate and postselection of the state $\ket{\downarrow}$. As we show below, for the measurement of the nonlocal parity operator $\mathcal{P}$, a system with $\mathcal{P} = +1 $ is mapped to the meter state $\ket{\psi_\mathcal{M}} = \ket{\downarrow}$, while a system with $\mathcal{P} = -1 $ is mapped to the meter state pointing in the opposite direction. Additionally, our erasure-based nonlocal measurement outcomes for a general system are projected to the subspace of some nonlocal observables, for example, the subspace of parity. We also show that our scheme can be applied to Bell measurements, having a crucial role in many quantum  protocols.

The experimental setup is shown in Fig.\,\ref{fig:setup}. Two photons, labeled as $a$ and $b$, are prepared in a polarization-momentum hyper-entangled state\,\cite{Kwiat1997} which is generated via the degenerate spontaneous parametric down conversion (SPDC). We take the polarization as the system $\mathcal{S}$ and use the correspondence $H(V)\leftrightarrow\uparrow(\downarrow)$ to spin-$\frac{1}{2}$ particles, where $H(V)$ stands for horizontal (vertical) polarization. The system's state can be prepared in any form by inserting HWP-QWP (which stand for half and quarter wave plates respectively) sets in each output arm. The photons' momentum degree of freedom (up and down paths labeled by $u$ and $d$, not to be confused with the aforementioned $\uparrow/\downarrow$ spin directions) is adopted as the ancilla. This structure guarantees that the photons' momenta are initialized in an EPR state $\ket{\psi_\mathcal{N}} = \frac{1}{\sqrt{2}}(\ket{ud} + \ket{du})$. For recording the final measurement results, we introduce another meter qubit $\mathcal{M}$ on Bob's side, left $(l)$ and right $(r)$ paths of photon-$b$, as shown in Fig.\,\ref{fig:setup}(a).

In Figs.\,\ref{fig:setup}(b) and (c), we show the operations on Alice's and Bob's side, respectively. For further clarification, we also draw the corresponding circuit diagrams for both of them. In our protocol, Alice firstly performs a C-NOT operation over the system and ancilla realized by BD1, where the photon's momentum is shifted conditionally on its polarization. To select the result $\sigma_z = -1$, a hole is used to block the up path and only collect the photons in the down path. Then Bob performs two C-NOT operations, one over the system and meter implemented by BD2, and the other over the ancilla and meter implemented by BD3. Finally, two HWPs are adopted with their optical axes oriented at $22.5^\circ$, as well as BD4 for implementing the quantum erasure of the ancilla. The photons in path-$d$ are collected by the fiber collimator after a hole. Four HWPs are inserted between BD2 and BD3 for guiding the photons to the $r$-mode, where the fiber collimator is located. To be more specific, when the optical axes of HWP-$l$s are oriented at $45^{\circ}$ while those of HWP-$r$s are oriented at $0^{\circ}$, the two $l$ beams are collected and others are blocked, as shown in Fig.\,\ref{fig:setup}(c). Conversely, the two $r$ beams are picked up.

\begin{table}[b]
\caption{\label{tab:ParityCheck} Coincidence counts for parity checks}
\begin{ruledtabular}
\begin{tabular}{clcccccccc}
\multicolumn{2}{c}{Channel} &\multicolumn{4}{c}{$l$}&\multicolumn{4}{c}{$r$}\\\cline{3-10}
\multicolumn{2}{c}{Basis} & HH  & HV  & VH  & VV & HH  & HV  & VH  & VV \\\hline
\multirow{2}{*}{$\mathcal{P}=+1$} & $|HH\rangle$  & 9192 & 17  & 23  & 0    & 11  & 23   & 10    & 0 \\
   & $|VV\rangle$  & 0    & 18  & 17  & 9405 & 0   & 6    & 21    & 8 \\
\multirow{2}{*}{$\mathcal{P}=-1$} & $|HV\rangle$  & 25   & 18  & 0   & 14   & 14  & 9258 & 0     & 18 \\
   & $|VH\rangle$  & 9    & 0   & 13  & 25   & 24  & 0    & 9412  & 15 
\end{tabular}
\end{ruledtabular}
\end{table}

\section{Parity check}
In our experiment, we first measured the system's parity $\mathcal{P}$. If the system is in the state $\ket{HH}$ or $\ket{VV}$, it is supposed to take the value $\mathcal{P} = +1$, and for states $\ket{HV}$ or $\ket{VH}$, it takes the value $\mathcal{P} = -1$. Table\,\ref{tab:ParityCheck} indicates the coincidence counts for the corresponding channel and polarization. It is clearly shown that, when the system's state is $\ket{HH}$ or $\ket{VV}$ (corresponding to $\mathcal{P} = +1$), photon-$b$ only comes out in channel $l$, and when the system is in state $\ket{HV}$ or $\ket{VH}$, only channel $r$ gives counts. The error rate is only around 0.2\%, which is mainly due to the non-perfect interference and the finite extinction ratio of the polarizer. Our results imply that the outcomes of the parity measurement of the system can be accurately revealed by the path state of photon-$b$.

\begin{figure}
   \centering
   \includegraphics[width =0.48\textwidth]{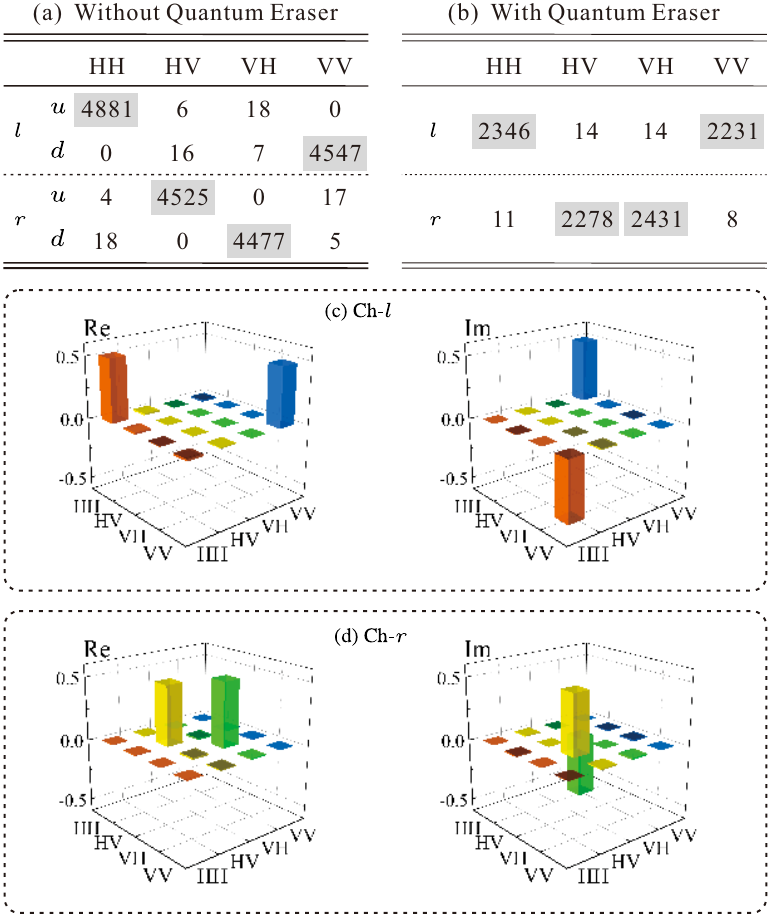}
   \caption{\textbf{Comparison of the measurement procedure with and without quantum erasure.} The coincidence counts for the two scenarios are shown in (a) and (b) respectively. In (c) and (d), the complete states, including the phase information in the corresponding output channel, are presented by the density matrices experimentally reconstructed, whose real and imaginary parts are given on the left and right, respectively.}
   \label{fig:QEraser}
\end{figure}

\section{Quantum erasure} 
One of the crucial steps in our method for implementing nonlocal measurement is the quantum erasure of local information. This is because we wish to infer the degenerate eigenvalue of the product of two operators, without knowing their local values. 
We show the erasure procedure in Figs.\,\ref{fig:QEraser}(a) and (b). Here the system is initialized in the state $\ket{\psi_\mathcal{S}} = \frac{1}{2}(\ket{HH}-i{VV}-i{HV}+{VH})$. The coincidence counts for each basis and channel before and after performing quantum eraser are compared. As shown in Fig.\,\ref{fig:QEraser}(a), immediately after Bob couples the meter $\mathcal{M}$ to both the system $\mathcal{S}$ and his part of the ancilla $\mathcal{N}$, the four basis elements, i.e., the eigenstates of the nonlocal observable $\sigma^A_z\sigma^B_z$, $\ket{HH}$, $\ket{HV}$, $\ket{VH}$, $\ket{VV}$ are directly coupled to the four path states, i.e., $\ket{lu}$, $\ket{ru}$, $\ket{rd}$, $\ket{ld}$, respectively. That is, the system's information can then be determined completely by measuring the path state. For coupling the measurement result of a nonlocal observable to a single pointer, we erase the redundant quantum information contained in the ancilla $\mathcal{N}$. Experimentally, we guide the two modes $\ket{u}$ and $\ket{d}$ to a beam splitter which acts as a Hadamard gate and collect photons only in the output port $\ket{d}$, as shown in Fig.\,\ref{fig:setup}(b). We show the coincidence counts in the corresponding channels for each basis in Fig.\,\ref{fig:QEraser}(b), which strengthens our conclusion, i.e., channel $l$ only contains the components $\ket{HH}$ and $\ket{VV}$ with even parity, and channel $r$ only contains the components $\ket{HV}$ and $\ket{VH}$ with odd parity, with a small error rate around 0.6\%.

\section{Subspace projection} 
We can also see the robustness of our experiment when the system is initialized in a general state (the same one as in the previous section), i.e., superposition of all the four bases. In this scenario, our setup can not only divide the photons into two parts according to the system's parity but also preserve the quantum coherence, which means our nonlocal measurement method is actually non-demolition and the system will be coherently projected onto the corresponding subspace according to the outcomes of the nonlocal measurement. We verify these results by performing additional polarization tomography on the system after the measurement procedure is finished, i.e., before the photons are entering the fiber collimators, we insert polarization analyzers (composed of QWP-HWP-PBS) on both Alice's and Bob's sides. The results are shown in Figs.\,\ref{fig:QEraser}(c) and (d). Compared to the theoretical expectations, the fidelities read $0.992\pm0.14$ and $0.984\pm0.007$, respectively. The non-degenerate imaginary parts of the density matrices clearly show that the coherence of the polarization states in the two parts is preserved in the nonlocal measurements, which is crucial for us to perform subsequent measurements. These sanity checks show that the erasure protocol works well, as theoretically planned, though up to now the state was separable and a local procedure could have succeeded as well. 

\section{Bell measurement} 
\begin{figure}
   \centering
   \includegraphics[width =0.45\textwidth]{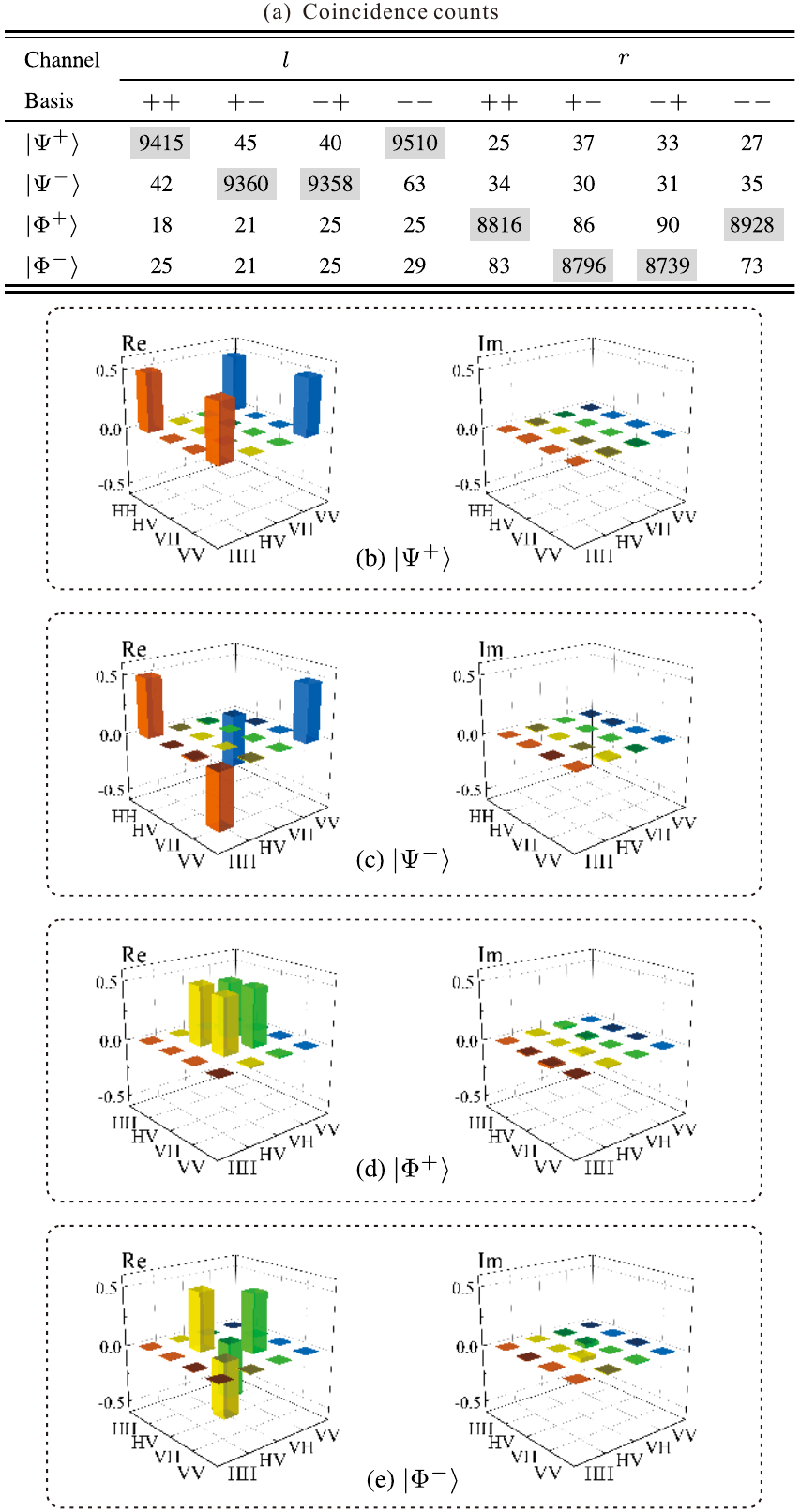}
   \caption{\textbf{Bell measurement.} (a) shows the coincidence counts in 10 seconds for each given channel and basis. (b)-(e) present the corresponding density matrices reconstructed from the correct output channel. The basis states $\ket{+}$\ and $\ket{-}$ stand for the eigenstates of $\sigma_x$ with eigenvalues $+1$ and $-1$, respectively.}
   \label{fig:Bell}
\end{figure}

Based on the above demonstration, we consequently realized a full Bell measurement of entangled pairs using the erasure-based nonlocal measurements. This was done without invoking a two-photon Hong-Ou-Mandel interference (such as demonstrated in\,\cite{Carsten2006,Li2016,Edamatsu2016}). Rather, we have just performed some simple projective measurements after the parity measurement. As\,\cite{Brodutch2016} implies, after picking out the Bell states $\ket{\Psi^{\pm}}=\frac{1}{\sqrt{2}}(\ket{HH}\pm \ket{VV})$ and  $\ket{\Phi^{\pm}}=\frac{1}{\sqrt{2}}(\ket{HV}\pm \ket{VH})$ via the path states $l$ and $r$, we just need to project onto the eigenstates of $\sigma_x$ for both photon-$a$ and photon-$b$, respectively, and then simply multiply the results of the local measurements. If $\sigma_x^A\otimes\sigma_x^B = +1$, it should be the state $\ket{\Psi^+}$ or $\ket{\Phi^+}$, and on the other hand if the product equals $-1$, the state should be $\ket{\Psi^-}$ or $\ket{\Phi^-}$. Combining the results of the two steps we can completely and deterministically distinguish between the four Bell states. The results are shown in more detail in Fig.\,\ref{fig:Bell}. We can see that $\ket{l,+1}, \ket{l,-1}, \ket{r,+1}, \ket{r,-1}$ stand for the Bell states $\ket{\Psi^+}$, $\ket{\Psi^-}$, $\ket{\Phi^+}$, $\ket{\Phi^-}$, respectively. We have also performed tomography for the purposes of verification and concluded that the density matrices coincide with our predictions very well as shown in Fig.\,\ref{fig:Bell}(b-e), where the fidelities compared to their theoretical expectation read $0.986\pm0.015$, $0.980\pm0.007$, $0.974\pm0.018$ and $0.983\pm0.006$, respectively. These results prove the validity and experimental applicability of the theoretical proposal in\,\cite{Brodutch2016}.

\section{Conclusions}
In conclusion, we have demonstrated an erasure-based scheme for performing nonlocal quantum measurements in a photonic system. The time evolution and measurements are performed on the photon's polarization degree of freedom. The outcomes of the nonlocal observable are directly given by a single local pointer. 
Our scheme is actually a nondemolition measurement of the nonlocal observable where quantum coherence is preserved during the process. As a consequence, our result can be extended to more complex protocols where further measurements are needed, for example, investigating causal roles and extracting nonlocal information in quantum networks and other distributed systems\,\cite{Ringbauer2016,Dressel2017,Manikandan2018}. In addition, we employed this scheme for performing a complete Bell measurement which is free of classical communication. The protocol is, and must be, probabilistic to preserve relativistic causality. Using this method one can realize any controlled-controlled-unitary gate and hence it could have many more applications for quantum information processing in nonlocal systems. Importantly, we have effectively generated a tripartite nonlocal interaction with negligible imperfections, which is very useful for photonic quantum computers.



\section{Acknowledgements}
Wei-Wei Pan and Xiao-Ye Xu contributed equally to this work. We wish to thank Aharon Brodutch for helpful comments and discussions. This work was supported by National Key Research and Development Program of China (Nos.\,2017YFA0304100, 2016YFA0302700), the National Natural Science Foundation of China (Nos.\,61327901, 11474267, 11774335, 61322506, 11821404), Key Research Program of Frontier Sciences, CAS (No.\,QYZDY-SSW-SLH003), Science Foundation of the CAS (No. ZDRW-XH-2019-1), the Fundamental Research Funds for the Central Universities (No.\,WK2470000026), the National Postdoctoral Program for Innovative Talents (No.\,BX201600146), China Postdoctoral Science Foundation (No.\,2017M612073), and the Anhui Initiative in Quantum Information Technologies (Grant No. AHY020100, AHY060300).

\bibliographystyle{apsrev4-1}
\bibliography{references}

\end{document}